\documentstyle[aps,twocolumn,epsfig]{revtex}

\newcommand{\infig}[2]{\begin{center}{\epsfig{file=#2,width=#1}}\end{center}} 

\begin{document}
\twocolumn[\hsize\textwidth\columnwidth\hsize
\csname @twocolumnfalse\endcsname

\title{Entanglement engineering of one-photon wavepackets using 
a single-atom source}
\author{K.M. Gheri$^{\rm (1)}$, C. Saavedra$^{\rm{(1,2)}}$, P. T\"{o}rm\"{a}$^{\rm (1)}$, 
J. I. Cirac$^{\rm (1,3)}$, and P. Zoller$^{\rm (1,3)}$}
\address{(1) Institut f\"{u}r Theoretische Physik, Universit\"{a}t Innsbruck,
Technikerstrasse 25/2, A-6020 Innsbruck, Austria \\
(2) Departamento de F\'{\i}sica,
Universidad de Concepci\'on, Casilla 4009, Concepci\'on, Chile\\
(3) ITP, University of California, Santa Barbara, CA 93106-4030}
\date{\today}

\maketitle

\begin{abstract} We propose a cavity-QED scheme for the controlled generation 
of sequences of entangled single-photon wavepackets. A
photon is created inside a cavity via an active medium, such as an atom,
and decays into the continuum of radiation modes outside the cavity.
%(coupled, for example to an optical fiber)
Subsequent
wavepackets generated in this way behave as independent logical
qubits. This and the possibility of producing maximally entangled
multi-qubit states suggest many applications in quantum communication.
\vskip0.5pc
Pacs number(s): 03.67.Hk, 03.67.-a
\end{abstract}

\vskip1.5pc]

Sources offering a great variety of entangled states are required for
the implementation of many quantum communication and computation
protocols\cite{DiV95,Be95}. With quantum communication~\cite{czkm97} in mind the
choice of photons as qubits is especially appropriate, since they can be
easily transfered over long distances. The standard source presently
used in the lab is parametric downconversion in a crystal
\cite{bpmewz97,bbmhp97}. It is a reliable source of entangled
twin-photons but the process is random and largely untailorable.
Moreover, in practice its capability of generating entanglement is limited to states
comprising only two photons. In this Letter we propose a scheme for the
controlled generation of many entangled photonic qubits. Our source of
entanglement produces a train of single-photon wavepackets which are
well resolved in time. This permits us to regard them as individual
qubits. In its most simple implementation the setup consists of a single
multilevel atom inside an optical resonator \cite{Turchette95,ion}. The
individual wavepackets are generated by applying an external laser
pulse to the atom prepared in a superposition state of
its internal states. The coupling of the atom to the resonator allows
the transfer of a single photon to the resonator and therefrom via
cavity decay to the continuum of radiation modes outside the resonator
(possibly coupled to an optical fiber). An
encoding of quantum information in the one-photon wave-packets could
either take place by identifying two orthogonal polarization states of
the single photon with logical ``0'' and ``1'', or by regarding the
absence of a photon as logical ``0'' while its presence would correspond
to logical ``1''.

Our scheme offers a twofold advantage over already existing sources of
entangled single-photon wavepackets such as down-conversion. 
 It provides excellent control over the instances in time
when a qubit is created as well as over the spectral composition of the
wavepacket. The qubits may thus be generated with a well defined {\em
tact frequency\/} and pulse shape. Moreover, repeated coherent 
recycling of the state of the atom after the generation of a photon wave packet 
gives rise to higher order entanglement between subsequent photons \cite{Hagley97}. 
In this regard our scheme generalizes and extends recent work on sources of 
single photon wavepackets, commonly referred to as 
{\em photon guns\/}~\cite{lk97} or turnstile devices~\cite{mn96} 
by allowing the generation of entangled multiphoton states. 
In particular, states such as the the three-particle GHZ state, and more 
generally $n$-qubit maximally entangled states (MES) can be generated. 
Since the individual wavepackets do not overlap in time, 
each can be sent to a different receiver node using simple classical 
gating operations. Such high-order entangled qubit states 
have immediate application in quantum cryptography 
\cite{bb84} and teleportation \cite{bbcjp93}, as well as in 
tests of non-locality and multiparticle interference \cite{multi}. 
Because entangled states of more than two qubits can be generated in a
straightforward manner, our scheme is especially useful for quantum
communication between many parties \cite{mn95}.

\begin{figure}[hbt]
\infig{7cm}{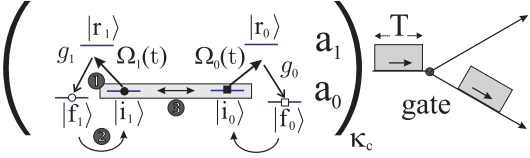}
\caption[]{
A single atom with six internal states interacts with two cavity 
modes of orthogonal polarization $a_0$, $a_1$.
 In a  Raman process
(step 1) 
an initial superposition state of levels $|i_0\rangle$ and $|i_1\rangle$ is 
transformed into an entangled cavity-atom state. Due to cavity leakage 
the photon will leave the cavity and produce a photon wave packet 
in the continuum modes outside the resonator. 
In step 2 the atom is 
recycled back to $|i_0\rangle$ and $|i_1\rangle$.
Between two photon generations levels $|i_0\rangle$ and $|i_1\rangle$ can be
coupled (step 3) to tailor the outgoing state.
}
\label{fig:system}
\end{figure}

Whilst the theory underlying our proposal can be formulated in a 
model-independent fashion it is more instructive to illustrate
the basic ideas using a specific model: we consider
a single atom or ion trapped inside a cavity \cite{Turchette95,ion}.
For the atom we assume a double three-level $\Lambda $ 
structure in the large detuning limit
as depicted  in Fig.~\ref{fig:system}.
The levels $|i_\alpha \rangle$ ($\alpha=0,1$) are coupled to the
upper levels $|r_\alpha \rangle$ via classical fields $\Omega_\alpha
(t)e^{-i(\omega _{\alpha} t+\phi_\alpha (t))}$, where $\omega _\alpha$ are 
the field center frequencies and the subscript refers to the two polarization states.
The external control parameters are the real 
amplitudes $\Omega_\alpha(t)$ and the phases $\phi_\alpha (t)$. The levels 
$|f_\alpha\rangle $ are coupled to the upper levels by the cavity modes 
$a_\alpha$ (common frequency $\omega _c$ but orthogonal polarization),
with coupling constants $g_\alpha$. The large detuning ($\delta$) assumption 
allows us to adiabatically eliminate the upper atomic levels. We are left 
with two two-level systems describable by generalized spin operators
$\sigma_{i_\alpha j_\alpha}=|i_\alpha\rangle\langle j_\alpha|$. 
The center frequencies of the external laser pulses fulfil the Raman
resonance condition. Note that any offsets can still be accommodated in
the phases $\phi_\alpha(t)$. The field outside the resonator is
described by a continuum of harmonic oscillators with creation and
annihilation operators $b_\alpha^{\dagger}(\omega )$, $b_\alpha(\omega
)$, respectively. The reservoir modes satisfy standard bosonic commutation
relations: $[b_\alpha(\omega ),b_\beta^{\dagger
}(\nu)]=\delta_{\alpha\beta}\delta (\omega -\nu )$. The coupling of the
cavity and reservoir modes is assumed to be flat around the cavity
resonance frequency and equal to
$\sqrt{\kappa _c/\pi }$ \cite{Gardiner}.
Hence the total system consists of three building blocks: the continuum 
outside the resonator, the cavity modes and the internal degrees of freedom 
of the atom inside the resonator. We switch to an
interaction picture with respect to the free dynamics of the compound system. This eliminates
the fast optical timescales from the dynamics and leaves us
with a simpler Hamiltonian:
\begin{eqnarray*}
H(t)&=& \sum_\alpha \biggl (i\sqrt{\frac{\kappa _c}{\pi}}
\int d\omega\left( a_\alpha b_\alpha^{\dagger}
(\bar{\omega} ) e^{i\omega t}- \rm{H.\ c.}
\right)\biggr ) + V(t),\\
V(t) &=& \sum_\alpha \biggl (\bar{\Omega}_\alpha^2(t)
\sigma_{i_\alpha i_\alpha}
+|\bar{g}_\alpha|^2a_\alpha^{\dagger }a_\alpha
\sigma_{f_\alpha f_\alpha} \biggr .\\
&+&\biggl . ir_\alpha(t)\left( e^{-i\phi_\alpha (t)}a_\alpha^{\dagger }
\sigma_{f_\alpha i_\alpha}
- e^{i\phi_\alpha (t)}
\sigma_{i_\alpha f_\alpha} a_\alpha\right ) \biggr ).
\end{eqnarray*}
We have introduced the following abbreviations 
$\bar{\Omega}_\alpha(t)=\Omega_\alpha(t)/2\sqrt{\delta}$, 
$\bar{g}_\alpha=g_\alpha/\sqrt{\delta}$,
$r_\alpha(t)=\bar{g}_\alpha\bar{\Omega}_\alpha (t)$ and 
$\bar{\omega}\equiv \omega+\omega_c$. The time and intensity dependent 
terms in $V(t)$ correspond to ac-Stark shifts arising from the adiabatic 
elimination of the upper atomic levels.

We assume an atom initially prepared in a superposition state
$|\varphi(0)\rangle_a=c_0|i_0\rangle+c_1|i_1\rangle$ which we wish
to map onto a superposition of continuum (reservoir) excitations.
The cavity and the reservoir modes are in their vacuum states.
Since the dynamics contain {\em
no\/} polarization mixing terms we may independently
consider those degrees of freedom corresponding to a single index
$\alpha$. We may thus work with a smaller system and intermittently drop
the index $\alpha$. The final state of the total system can be obtained
using the superposition principle and issuing each partial solution with
the appropriate probability amplitude $c_\alpha$.
As a starting point we will discuss briefly the generation of single
photon wavepackets entangled with the atom.
Since there can be at most a single excitation 
transfered to the continuum we are led to the following ansatz for the
state of the total system:
\begin{equation}
|\psi (t)\rangle =|\varphi (t)\rangle_{\rm{ac}} | 0 \rangle_{\rm{r}}
+\int d\omega |\varphi _{\bar{\omega} }(t)\rangle_{\rm{ac}} 
|\rm{1}_{\bar{\omega}} \rangle _r.   
\end{equation}
Here $|\varphi _{\bar{\omega}} (t)\rangle _{\rm{ac}}$ and $|\varphi
(t)\rangle_{\rm{ac}}$ denote atom-cavity states with and without 
the transfer of a photon having taken place into the reservoir mode with 
frequency $\bar{\omega} $, respectively.
Note that $|\varphi (t)\rangle_{\rm{ac}}$ 
describes the atom-cavity state before the photon has been lost to the 
reservoir:
% it is thus given by 
\[
|\varphi (t)\rangle _{\rm{ac}}=C_{i}(t)e^{-i\theta (t)}|i\rangle
|0\rangle_{\rm{c}}
+C_{f}(t)e^{-i|\bar{g}|^2t}|f\rangle |1\rangle_{\rm{c}} ,
\]
where $\theta (t)=\int_0^tdt^{\prime }\bar{\Omega}^2 (t^{\prime })$. 
Applying $a$ to this state projects the coupled atom-cavity system into 
the state $|f\rangle|0\rangle_c$ which is not coupled by $V$. We thus find: 
\begin{equation}
|\varphi _{\bar{\omega} }(t)\rangle_{\rm{ac}}=\sqrt{\frac{\kappa _c}\pi }%
\int_0^tdt^{\prime }e^{i\omega 
t^{\prime }}a|\varphi (t^{\prime })\rangle_{\rm{ac}}. \label{eq:phi1}
\end{equation}
If we insert this expression into the equation for 
$|\varphi (t)\rangle _{\rm{ac}}$ and perform the Markov approximation we 
arrive at a simple closed equation:
\begin{equation}
{|\dot{\varphi} (t)\rangle_{\rm{ac}} } =-(\kappa_c a^\dagger a + iV(t))
|\varphi (t)\rangle_{\rm{ac}}.\label{eq:Markov}
\end{equation} 
This now permits us to specify the sought evolution equations for the 
coefficients in the ansatz for $|\varphi (t)\rangle_{\rm{ac}}$:
\begin{eqnarray*}
\dot{C}_{i}(t) &=&
-r(t)C_{f}(t)\exp(i\theta_c (t)), \\
C_{f}(t) 
&=&\int_0^tdt^{\prime }r(t^{\prime })C_{i}(t^{\prime})
\exp(-(\kappa_c(t-t')+i\theta_c (t'))),
\end{eqnarray*}
where $\theta_c(t)=\theta(t)+\phi(t)-|\bar{g}|^2t$.
In the limit of an overdamped cavity the integral will get a non-zero 
contribution only from those times $t'$ which are close to $t$ on the 
scale of the cavity lifetime $\kappa _c^{-1}$. To good approximation 
it thus holds that
\begin{eqnarray*}
C_{f}(t) &\simeq &\frac{r(t)}{\kappa _c}
e^{-\mu (t)-i\theta_c (t)},\quad C_{i}(t) =e^{-\mu (t)}  ,
\end{eqnarray*}
where $\dot{\mu} (t)=r^2(t)/\kappa _c$, and $\mu(0)=0$.
The actual object of interest is the state of the continuum of radiation modes
outside the cavity. We thus 
insert the above result for $C_{f}$ into Eq.~(\ref{eq:phi1}) and find:
\begin{eqnarray}
|\varphi _{\bar{\omega} }(t)\rangle &=&\sqrt{\frac{\kappa _c}\pi }%
\int_0^tdt^{\prime }e^{i\omega 
t^{\prime }} \frac{r(t')}{\kappa_c}e^{-\mu (t')-i(\theta(t')+
\phi(t'))}|f\rangle|0\rangle_{\rm{c}}\nonumber\\
&\equiv&G(\omega,t)\sigma_{fi}|i\rangle|0\rangle_{\rm{c}} .
\label{eq:cont}
\end{eqnarray}

Eq.~(\ref{eq:cont}) indicates a direct mapping of the initial
atomic state to the final one accompanied by the creation of a
wavepacket with spectral envelope $G(\omega,t)$, cf.\ step 1 in
Fig.~\ref{fig:system}.  To make
the scheme practical two constraints have to be kept in mind. First of
all we would like to implement an efficient transfer of the photon to
the continuum. Secondly, we are only interested in pulse sequences
that terminate after a finite time $T\gg\kappa_c^{-1}$. This would be
warranted if at the time $t=T$ the atom is with near certainty in its
internal state $|f\rangle$, i.e., iff $\mu(T)\gg
1$. Recalling the definition of $\mu(t)$ this sets for any given
pulse duration $T$ a lower bound for the minimum size of the pulse
area of the applied laser field. Under this assumption the
total system state for times $t>T$ is given by:
\begin{equation}
|\psi(t)\rangle= \biggl (\sum_\alpha c_\alpha B_\alpha^\dagger(0,T)
|f_\alpha\rangle\biggr )  
|0\rangle_r |0\rangle_{\rm{c}_0}|0\rangle_{\rm{c}_1},
\end{equation}
where $B_\alpha^\dagger(t_j,T)$ is the creation operator of a one-photon 
wavepacket with logical or polarization state $\alpha$ within the time 
window from $t_j$ to $t_j+T$:
\begin{equation}
B_\alpha^\dagger(t_j,T)= 
\int d\omega e^{i\omega t_j} G_\alpha(\omega,T)b_\alpha^\dagger(\bar{\omega}).
\label{eq:boper}
\end{equation}
Note that the spectral envelope now carries a subscript as the system 
parameters need not be the same for each of the effective two-level systems 
we use to implement the mapping. In brief, we have shown how one can transform 
an initial atomic superposition state into an entangled atom-continuum state. 
Had we used only a single $\Lambda$ system we would have recovered the 
{\em photon gun\/} \cite{lk97}, a tailorable simple single-photon source. 
The novel aspect here is the residual entanglement between the internal atomic 
state and the polarization state of the outgoing wavepacket. We may harness 
this to create a sequence of entangled one-photon wavepackets which are well 
resolved in time. Let us introduce the following 
abbreviation for a one-photon wavepacket with polarization $\alpha$ that has 
been generated in the $j$-th generation sequence:
$|\alpha\rangle_j=B_\alpha^\dagger(t_{j-1},T)|0\rangle_r$ (with $t_0=0$).
The state after the first sequence in more compact form reads:
\begin{equation}
|\psi(t)\rangle=(c_0|0\rangle_1|f_0\rangle+c_1|1\rangle_1|f_1\rangle)
|0\rangle_{\rm{c}_0}|0\rangle_{\rm{c}_1}.
\label{after1}
\end{equation}
Suppose we apply a further pulse sequence which recycles the atom back to its 
initial state, i.e., $|f_\alpha\rangle\rightarrow|i_\alpha\rangle$. Then at 
a time $t_1>T$ we reinitiate the same pulse sequence that we have already used 
previously. It is plausible that the wavepackets already generated have in the 
meantime propagated far away from the cavity and thus cannot influence the 
renewed generation sequence. Going through the same procedure again we obtain 
for $t>t_1+T$:
\begin{equation}
|\psi(t)\rangle=(c_0|0\rangle_2|0\rangle_1|f_0\rangle+
c_1|1\rangle_2|1\rangle_1|f_1\rangle)
|0\rangle_{\rm{c}_0}|0\rangle_{\rm{c}_1}.
\label{eq:step2}
\end{equation}
The residual entanglement with the generating
system can eventually be broken up by making a measurement of the
internal atomic state in an appropriate basis, e.g.,
$|f_0\rangle\pm|f_1\rangle$. For the state in
Eq.~(\ref{eq:step2}) the resulting reservoir state would be one of two
states $c_0|0\rangle_1|0\rangle_2\pm c_1|1\rangle_1|1\rangle_2$.
Repeating the generation process $n$-times followed by a final state
measurement we produce an $n$-photon wave packet.
Note that the description of the
reservoir state in terms of products of one-photon wavepackets implies that the
wavepackets can be regarded as independent quantum entities. It has to be 
emphasized that such a description is only possible because of the vanishing 
temporal overlap of the individual outgoing wavepackets. Actually components of
the reservoir state are given by products of the operators 
$B_\alpha^\dagger(t_j,T)$ applied to the multimode vacuum $|0\rangle_r$. By 
construction operators belonging to different sequences, cf.\ Eq.~(\ref{eq:boper}), 
however, commute to good approximation:
\begin{eqnarray}
[B_\alpha(t_k,T),B_\beta^\dagger(t_j,T)]&=&\delta_{\alpha\beta}
\int d\omega e^{i\omega (t_j-t_k)} |G_\alpha(\omega,T)|^2\nonumber\\
&\approx& \delta_{\alpha\beta}\delta_{jk}.
\end{eqnarray}
Formally, we may thus regard each creation operator as acting on its own
vacuum state. Physically, this corresponds to the fact that each
wavepacket is contained within its private time window of duration $T$
or a box of length $cT$ with no overlap between successively generated
packets, cf.\ Fig.~\ref{fig:system}. We have numerically checked the
factorization assumption for a two-photon wavepacket modeling the
reservoir by a discrete set of 1024 ``continuum''-modes (amounting to
more than $10^6$ reservoir states) embedded in a frequency window of
width $40$ $\kappa_c$. We found that both the Markovian approximation
used in Eq.~(\ref{eq:Markov}) and the factorization assumption for the
two-photon spectral density are excellent with relative errors of the
order of $10^{-3}$.

For any source of entanglement it is essential to fathom what the accessible class of
states is.
%it is capable of generating.
In general, a state in a basis
spanned by $n$-qubits is defined by $N=2^{n+1}-2$ 
independent coefficients. In our specific model
the states can be tailored by coupling the levels $|f_{\alpha}\rangle$
by a microwave/Raman pulse inbetween the qubit generation sequences. For
example, this transforms Eq.~(\ref{after1}) into $|\psi\rangle = (c_0 d_0
|0\rangle_1 |f_0\rangle + c_0 d_1 |0\rangle_1 |f_1\rangle + c_1 d^*_0
|1\rangle_1 |f_1\rangle - c_1 d^*_1 |1\rangle_1 |f_0\rangle)
|0\rangle_{c_0} |0\rangle_{c_1}$. The coefficients $d_\alpha$ can be
chosen at will. With each applied pulse two independent parameters are
introduced. For $n$ qubits we have thus $2n$ free parameters at hand for
the purpose of state engineering. Since this is much less than $N$, 
only a restricted subclass of states can be created. However, we 
emphasize that the accessible class of states includes many useful 
and interesting states. For example MES such as the four Bell states, 
the GHZ-state $(|000\rangle+|111\rangle )/\sqrt{2}$ 
and its higher dimensional counterparts
[$(|s_1,s_2,...,s_n\rangle+|1-s_1,1-s_2,...,1-s_n\rangle )/\sqrt{2}$,
$s_i=0,1$] can easily be produced.
 
The maximum number $n$ of entangled photon wavepackets (qubits) our scheme can generate is
limited by decoherence.  Relevant sources of decoherence are:  (i) laser
phase and amplitude fluctuations; (ii) spontaneous emission during the
atomic transfer; (iii) absorption in the cavity mirrors; (iv) atomic motion.
The chosen configuration minimizes these effects.  
Stabilization of laser phase fluctuations below 1kHz represents a technical
challenge.  In the present scheme, the state produced after each cycle only depends on
the phase difference between the laser beams driving both transitions in Fig.\ 1.
% (the sum of the phases becomes a global phase that factorizes out). 
When these two laser beams
are derived from the same source the fluctuations in the phase difference are effectively
suppressed. Amplitude fluctuations cause a 
distortion of the pulse form and lead to incomplete population transfer. An estimate
gives that $n\ll I/\Delta I\sim 10^4$, where $\Delta I/I$ are the relative
intensity fluctuations.
Spontaneous emission from the auxiliary levels $|r\rangle$ at rate $\Gamma$
is quenched by choosing a
large detuning $|\delta|$ which leads to an effective decay rate 
$\Gamma_{\rm eff}=\Gamma \Omega^2/4\delta^2$ with $\Omega={\rm max} (\Omega_\alpha,g_\alpha)$.
For a peak frequency $\Omega_0=55$ Mhz and a Gaussian pulse shape for the classical
field $\Omega (t)$, $\delta=1.5$ GHz, $g=55$ Mhz, and $\kappa_c=50$ Mhz
one-photon pulse durations of around 10 cavity lifetimes are possible.
Recycling and reinitialization of the medium included a
conservative estimate would yield a generation rate of around $1$ MHz.
For $\Gamma=5$MHz the probability of spontaneous emission per cycle
is $<10^{-3}$. Photon absorption in the cavity mirrors is an essential effect
for high--Q optical cavities. In general, it leads to two types of errors:
Photon absorption and concomitant destruction of the entanglement.
These errors are evaded by postselection through
discarding sequences with a number of detected photons smaller than $n$.
State distortion~\cite{PellHid} can occur even in the absence of loss of a photon according to:
%if a single photon is absorbed, the whole entangled state will be destroyed;
%on the other hand, even if no photon is lost, the generated state will be
%distorted according to
\begin{eqnarray}
\label{error}
|\Psi\rangle &=& \sum_{x\in\{0,1\}^n} q_x |x\rangle \rightarrow 
%|\Psi'\rangle &\propto& 
\sum_{x\in\{0,1\}^n} q_x e^{-(\kappa_1-\kappa_0)T n_1(x)}|x\rangle
\end{eqnarray}
where the $x$ are binary representations of different photon states,
$n_1(x)$ is the number of ones contained in $x$, $\kappa_{0,1}$ are the
loss rate for modes $0$ and $1$, respectively.
% and $\tau$ is the cycle time. 
Errors as in Eq.\ (\ref{error}) vanish for
a cavity with equal absorption rates for both polarizations, i.e., $\kappa_0\simeq \kappa_1$.
%
%The modifications of the atomic motion due to the recoils experienced
%in the photon absorption and emission sequences will lead to an entanglement 
%of the state of the system and
%the motional state which amounts to decoherence. 
%Due to the spatial variations of the cavity modes and laser beams, at each position
%where the atomic wavepacket is significant, there will be an effective Rabi
%frequency; therefore, a complete cycle at a given atomic position will be incomplete
%for another position, which amounts to leaving population behind in the
%wrong levels. 
The optimal way to suppress fluctuations induced by the motion of the atom
is to place the atom at an antinode of both the cavity modes and
the laser beam (which have to be in standing wave configuration), where the
effect of spatial variations is minimum and operate
in the Lamb--Dicke regime~\cite{LDR}.
% where the size of the atomic wavepacket $L$ is smaller
%than the optical wavelength $\eta=L/\lambdabar\ll 1$. In this case, 
%one can easily determine that the population left in the wrong level is
%$\eta^4/4$, which puts a limit on the number of photon pulses $n\ll 4/\eta^4$.
Finally, there might be systematic and random errors in the adjustment of 
the laser pulses used in the recycling reinitialization.
\begin{figure}[bht]
\infig{4cm}{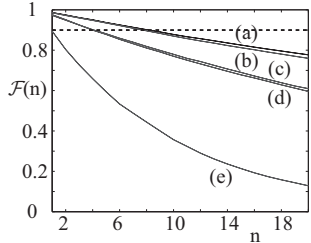}
\caption[]{Ensemble averaged fidelity as a function of the number of qubits $n$. 
Curves (a),(c), and (e) assume $\delta_m=0$ and
$\epsilon_m=0.0125,0.025,0.1$, respectively. Curves (b) and (d) are the same as (a) and (c) with 
$\delta_m=0.05$.}\label{fig:fidelity}
\end{figure}

\noindent To assess these effects we assume the following imperfect mapping in
each of the generation sequences:  $|f_\alpha\rangle \rightarrow A_\alpha
|i_\alpha\rangle + B_\alpha |f_\alpha\rangle$, where
$A_\alpha=(1-\epsilon_\alpha)\exp(i\delta_\alpha)$.  In Fig.\ 2 we plot the
fidelity ${\cal F}(n)$ of an $n$-qubit MES produced by a source which is ideal
exept that the magnitude and phase of $\epsilon_\alpha$ are evenly distributed
over a range of $[0,\epsilon_m]$ and $[-\epsilon_m,\epsilon_m]\pi$,
respectively, and the dephasing angle is evenly distributed over
$[-\delta_m,\delta_m]\pi$.  Fig.\ 2 shows that the process is rather robust
against global dephasing ($\delta_\alpha$) but that the correct timing of the
$\pi$-pulses is critical \cite{estimate}.  From curve (a) we gather that for
errors in the 2\% range approximately 10 qubits can be created with a fidelity
of 90\%.  

We have presented a CQED-based source for the controlled generation of
entangled $n$-qubit states where the individual qubits are nonoverlapping
one-photon wavepackets.  Our model seems experimentally feasible with
state-of-the-art equipment and could form the experimental basis for multi-party
communication in future quantum networks. The theory presented can be easily
adapted to other implementations (e.g. quantum dots or single atoms embedded
in a host material \cite{mn96}) which may emerge in the course of time as quantum systems
with long coherence times.

{\it Acknowledgements} 
%We thank H. Weinfurter for fruitful discussions.
This work was supported by the Austrian Research Foundation under 
grant no. S06514-TEC and the European TMR network ERB-FMRX-CT96-0087.
C.S. thanks Fundacion Andes for support.


\begin{references}
\bibitem{LDR}Here the size of the atomic wavepacket $L$ is smaller
than the optical wavelength $\eta=L/\lambdabar\ll 1$ and the population left behind 
in the wrong level is
$\eta^4/4$, which puts a limit on the number of photon pulses $n\ll 4/\eta^4$.
\end{references}
\end{document}